\documentclass[]{aa}
\usepackage[utf8]{inputenc}
\usepackage{graphicx}
\usepackage[colorlinks,linkcolor={blue},citecolor={blue},urlcolor={red}]{hyperref}
\usepackage{url}
\usepackage{natbib}
\usepackage{xspace}
\usepackage{amsmath}
\usepackage{multicol}
\usepackage{multirow}
\usepackage{makecell}

\def\apjs{ApJS}

\def\beq{\begin{equation}}
\def\eeq{\end{equation}}
\def\bey{\begin{eqnarray}}
\def\eey{\end{eqnarray}}

\def\Mpc{\,{\rm Mpc}}
\def\mpc{\, h^{-1}{\rm {Mpc}}}

\def\kms{\,{\rm {km\, s^{-1}}}}

\def\msun{\, h^{-1}{\rm M_\odot}}

\def\gs{\mathrel{\raise1.16pt\hbox{$>$}\kern-7.0pt
\lower3.06pt\hbox{{$\scriptstyle \sim$}}}}
\def\ls{\mathrel{\raise1.16pt\hbox{$<$}\kern-7.0pt
\lower3.06pt\hbox{{$\scriptstyle \sim$}}}}
\def\gtsima{\, {\buildrel > \over \sim} \,}
\def\ltsima{\, {\buildrel < \over \sim} \,}
\def\prosima{\, {\buildrel \propto \over \sim} \,}
\def\gsim{\lower.5ex\hbox{\gtsima}}
\def\lsim{\lower.5ex\hbox{\ltsima}}
\def\simgt{\lower.5ex\hbox{\gtsima}}
\def\simlt{\lower.5ex\hbox{\ltsima}}
\def\simpr{\lower.5ex\hbox{\prosima}}

\title{Anisotropy and characteristic scales in halo density gradient profiles}
\author{Xiaoyu Wang\inst{1,2}, Huiyuan Wang\inst{1,2}, H.J. Mo\inst{3} }
\institute{Key Laboratory for Research in Galaxies and Cosmology, Department of Astronomy, University of Science and Technology of China, Hefei, Anhui 230026, China; wambyybz@mail.ustc.edu.cn, whywang@ustc.edu.cn
\and School of Astronomy and Space Science, University of Science and Technology of China, Hefei 230026, China
\and Department of Astronomy, University of Massachusetts, Amherst MA 01003-9305, USA}
\date{November 2021}

\begin{document}
\abstract{
We use a large $N$-body simulation to study the characteristic scales in the density 
gradient profiles in and around halos with masses ranging from $10^{12}$ to 
$10^{15}\msun$. We investigate the profiles separately along the major ($T_1$) and minor 
($T_3$) axes of the local tidal tensor and how the characteristic scales  
depend on halo mass, formation time, and environment. We find two kinds of 
prominent characteristic features in the gradient profiles, a deep `valley' and 
a prominent `peak'. We use the Gaussian Process Regression to fit the gradient 
profiles and identify the local extrema to determine the scales associate with these features. 
Around the valley, we identify three types of distinct local minima, corresponding to 
caustics of particles orbiting around halos. The appearance and depth of the three 
caustics depend significantly on the direction defined by the local tidal field, 
formation time and environment of halos. The first caustic is located at a 
radius $r>0.8R_{200}$, corresponding to the splashback feature, and is 
dominated by particles at their first apocenter after infall. The second 
and third caustics, around $0.6R_{200}$ and $0.4R_{200}$ respectively,  
can be determined reliably only for old halos. 
%The radii of the three caustics are consistent with the prediction of self-similar gravitational collapse.
The first caustic is always 
the most prominent feature along $T_3$, but may not be the case along 
$T_1$ or in azimuthally-averaged profiles, suggesting that caution must 
be taken when using averaged profiles to investigate the splashback radius.  
We find that the splashback feature is approximately isotropic when 
proper separations are made between the first and the other caustics. 
We also identify a peak feature located at $\sim 2.5R_{200}$ in the 
density gradient profile. This feature is the most prominent along $T_1$
and is produced by mass accumulations from the structure outside halos.
We also discuss the origins of these features and their observational 
implications.
}

\maketitle

\section{Introduction}

In the standard cold dark matter paradigm of structure formation, dark matter halos 
are the building blocks of the cosmic web. Baryonic matter follows the collapse of dark matter 
and cools at the center of the halo to form galaxies \citep{Rees77,White78,Fall80,Mo10}. 
Understanding the properties and evolution of dark matter halos thus 
not only form the basis to model and characterize the large-scale structure of 
the universe, but also to model and understand galaxy formation. 
One important first step in understanding the halo population
is to discover salient properties related to their formation in the cosmic web.

In the simplest description, a dark matter halo is approximated by a 
sphere, within which the average mass density is a constant multiple ($\Delta$) of 
some reference density, $\rho_{\rm ref}$. The halo mass, $M_{\Delta}$, is 
defined as the enclosed mass, 
\begin{equation}
\label{eq_Mhalo}
\begin{aligned}
M_{\Delta}&=\frac{4}{3}\pi R^{3}_{\Delta} \Delta\rho_{\rm ref},
\end{aligned}
\end{equation}
where $R_{\Delta}$ is the radius of the sphere, or halo radius.
The reference density is generally taken as the critical or mean density of 
the universe, and the multiplier is chosen to be 200, roughly the value 
predicted by the virial theorem applied to the collapse of a spherical top-hat
perturbation in an expanding universe \citep{Gunn72,Gunn77,Peebles80,Mo10}. 
The halo radius is, therefore, also referred to as the virial radius of a halo. 
However, this approximate description cannot fully account for halo formation. 
For example, splashback substructures, which are physically associated with their host 
halo, are considered as isolated, independent halos in this description
\citep[e.g.][]{Ludlow2009ApJ, WangH09}. In addition, small halos can 
start to lose their mass long before they fall into the virial radius of a massive 
halo \citep{Behroozi14,Pearrubia17}, suggesting that the zone of influence of 
a halo may be bigger than that given by the virial radius. 
All these indicate that the density field outside $R_{\Delta}$ may also 
be physically connected to halos \citep[see e.g.][]{Prada06,Hayashi08,Oguri11,Diemer14,Trevisan17}
and should be characterized. Another potential problem with the traditional 
definition is that the growth of a halo based on the definition may sometimes be
un-physical. Indeed, based on the definition of Eq. \ref{eq_Mhalo}, halos 
that have ceased mass accretion can still have their $M_{\Delta}$ increase  
with time because of the increase in the halo boundary due to the decrease 
of $\rho_{\rm ref}$ \citep{Cuesta08,Diemer13,More15}.

 Great amounts of effort have been made to investigate the density profile 
in and around halos, in the hope of discovering some characteristic 
features and scales that reflect physical processes underlying halo formation
in the cosmic web. One of the most studied features is the so-called splashback radius, 
which was identified as the radius in the spherically 
averaged density where the gradient of the density profile is  
the most negative \citep[e.g.][]{Adhikari14, Adhikari16, Diemer14, More15, More16}. 
Subsequently, \cite{Mansfield17} measured the density field along different 
directions around a single dark halo and identified a shell-like structure 
by connecting the steepest points of the gradient profile in different 
directions. The splashback radius for massive halos is found to decline 
with increasing halo mass \citep[e.g.,][]{More15,Diemer17,O'Neil22} and 
with increasing mass accretion rate 
\citep[e.g.,][]{More15,Diemer17,Diemer20b,Contigiani21,O'Neil21}.
More recently, attempts have also been made to use hydro-simulations 
to study whether or not baryonic physics affects this characteristic scale
and whether or not baryons and stars follow the same trends as seen in dark 
matter \citep{Deason20,O'Neil21,O'Neil22}.

By tracking particles as they approach a halo and recording the time and location 
of their subsequent motion, \cite{Diemer17} and \cite{Diemer20b} found that
the splashback radius can roughly separate the in-falling matter from the matter 
that orbits around the halo \citep[see also][]{Adhikari14,Deason20,Sugiura20,Diemer21}. 
This suggests that the splashback feature is the result of a caustic produced  
by particles at their first apocenter after infall 
\citep[see e.g.][]{Diemand08,Vogelsberger09,Vogelsberger11,Diemer14,Adhikari14}. 
Higher-order caustics also seem to be present in simulated halos, although 
they are usually weak \citep[see e.g.][]{Diemand08,Deason20}.
%{\bf 
Caustics are also studied in details in the context of self-similar gravitational collapse \citep[e.g.][]{Fillmore84, Bertschinger85a, Mohayaee2006, Zukin2010}.
%} 

There are also efforts to define the boundary of a halo based on 
other considerations. For example, \cite{Fong21} defined a 
``characteristic depletion radius'' using the position of the minimum 
of the halo bias parameter as a function of radius.
\cite{Aung21a} proposed an ``edge radius'' to define the halo boundary based 
on the phase-space structure around halos. 

The presence of these characteristic 
scales in simulations have in turn inspired investigations to connect 
these scales with theoretical models and with observations.  
For example, \cite{Adhikari18} and \cite{Contigiani19a} used the splashback radius to 
distinguish models of modified gravity; %{\bf 
\cite{Gavazzi2006} and \cite{Mohayaee2008} suggested that the caustics can be used to detect dark matter;
%} 
\cite{Garcia2021MNRAS} and \cite{Fong21}
considered the implications of these scales in halo models of the large-scale 
structure; 
the connection with galaxy evolution was investigated by 
\cite{Deason20}\citep[see also][]{Dacunha22} and the link to accretion shocks by 
\cite{Anbajagane21}\citep[see also][]{Baxter21,Aung21b,Zhang21}.
Using projected galaxy number density profiles around redMaPPER galaxy clusters, 
\citep[][]{Rykoff14} and \cite{More16} measured splashback features associated with 
massive halos. They found a scale which is smaller than that expected from 
numerical simulations, possibly because of selection effects in the observational data
\citep{Zu17,Busch17,Chang18,Sunayama19,Murata20}. More recently,
galaxy clusters selected via the Sunyaev-Zel'dovich effect 
\citep[SZ,][]{Sunyaev72} have been utilized in such analyses to 
minimize the risk of spurious correlations between the splashback radius 
and cluster selection \citep{Shin19,Zurcher19,Adhikari21,Shin21}. 
%{\color{red} Need to say some of the finding here!}  
Attempts have also been made using a small sample of X-ray selected 
massive clusters \citep[e.g.][]{Umetsu17, Contigiani19b},  
spectroscopically confirmed galaxy members of massive clusters 
\citep[e.g.][]{Bianconi21}, and weak gravitational lensing \citep[e.g.][]{Umetsu17,Chang18,Contigiani19b,Contigiani21,Shin19,Shin21}.
%{\color{red} Need to say some of the finding here!} 
The splashback radii obtained from these observational data  
are broadly consistent with simulations, although the uncertainties 
in current data are quite large. \cite{Fong22} made a first attempt to 
measure the depletion radius using weak gravitational lensing data, 
and found that their results are consistent with those obtained from simulations. 

There is growing evidence that the characteristic scales may be anisotropic. 
\cite{Mansfield17} found a non-spherical splashback shell, the orientation of which 
is aligned with the mass distribution in the inner region of the halo. 
\cite{Contigiani21} found that the depth of the splashback feature in a cluster correlates 
with the direction of the filament containing the cluster and with the orientation of 
the brightest cluster galaxy. \cite{Deason20} found that the position of 
the splashback feature varies with the position angle relative to the 
neighboring structure. Such anisotropy is expected theoretically,   
as dark matter halos are found to be better described by ellipsoidal 
models \citep{Jing2002} and halo orientation is strongly aligned with the large-scale structure 
\citep[e.g.,][]{Hahn2007MNRAS, WangH11,Tempel13,Libeskind13, Chen16}.
However, a detailed analysis is still necessary to understand and characterize   
the anisotropy in the characteristic scales and its alignment with the  
large-scale structure. 

Most of the investigations so far have been focused on massive halos or halos 
with high accretion rate, and our knowledge about lower-mass and old halos 
remains relatively poor \citep[see e.g.][]{Mansfield17}. There are indications that 
halos of lower masses may have characteristic scales that are 
different from those in massive halos. For example, \cite{Deason20} 
found two caustic features in halos with masses similar to the 
Local Group \citep[see also][]{Diemer14,Adhikari14}; 
\cite{Fong21} found a transitional change in the ratio between the depletion 
radius and splashback radius as halo mass increases; there is also 
evidence that the formation of caustic depends on halo assembly \citep[e.g.][]{Diemand08}.
It is thus important to use a halo sample covering a large range in both mass and assembly 
history to fully understand the dependence on halo properties. 

In this paper, we use a large $N-$body simulation, the ELUCID
\citep{WangH16}, to investigate the characteristic scales of the 
density gradient profiles in different directions defined by the local 
tidal tensor, for halos with masses ranging from $10^{12}$ to $10^{15}\msun$. 
The rest of the paper is structured as follows. In Section \ref{sec_sam}, we briefly 
describe the ELUCID simulation, our halo catalog, the `halo tidal field', 
and our method to calculate and fit the halo density and gradient profiles. 
In Section \ref{sec_iden}, we show the halo profiles and check the performance of 
our fitting procedure. We then identify the characteristic scales and study their 
anisotropy and dependence on halo mass, assembly history and environment.
We also use the phase space density to examine signatures of the 
characteristic scales in phase-space. We summarize our results in 
Section \ref{sec_sum}.

\section{Data and Methods of Analysis}
\label{sec_sam}

\subsection{The simulation}
\label{sec_sim}

Our analysis is based on the ELUCID simulation \citep{WangH2014,WangH16}, 
which is a dark matter only, constrained simulation run with $3072^{3}$ dark matter 
particles (each with a mass of $3.088\times10^{8}\msun$) in a cubic box of 
$L_{\rm box}$ = 500$h^{-1}$ Mpc on each side. The simulation was run with L-GADGET, 
a memory-optimized version of GADGET2 \citep{Springel05}. The cosmological parameters 
adopted in the simulation are consistent with the WMAP5 \citep{Dunkley09} cosmology: 
density parameters in dark energy $\Omega_{\Lambda,0}$ = 0.742, 
in matter $\Omega_{m,0}$ = 0.258, and in baryons $\Omega_{b,0}$ = 0.044; 
Hubble's constant $h= H_0/100\kms\Mpc = 0.72$; the amplitude of density 
fluctuations $\sigma_8=0.8$; the index of initial perturbation power spectrum $n_s=0.96$. 
The simulation is run from redshift $z = 100$ to $z = 0$. Outputs are made at 100 snapshots, 
from $z = 18.4$ to $z = 0$, equally spaced in the logarithm of the expansion factor.

\subsection{Dark matter halos}
\label{sec_halo}

In the ELUCID simulation, dark matter halos were identified using a friends-of-friends (FOF) 
algorithm with a linking length equal to 0.2 times the mean particle separation 
\citep{Davis85}. Only halos containing at least 20 particles are identified. 
We used the SUBFIND algorithm \citep{Springel01} to identify subhalos in each FOF halo. 
This is a method to decompose a given FOF halo into a group of subhalos by using the 
local overdensity and an unbinding algorithm. The largest substructure is called the main halo. 

The halo mass, $M_{200}$, for a main halo is defined using Eq. \ref{eq_Mhalo} 
with $\Delta=200$ and $\rho_{\rm ref}$ equal to the cosmic mean density. 
It is thus the mass contained in the spherical region of radius $R_{200}$, 
centered on the most bound particle of the main halo, and within which the mean mass 
density is equal to 200 times of the cosmic mean density. \cite{Diemer14} found 
that $R_{200}$ is a more natural choice to scale the density profile 
at large radii than other definitions. They also found that the density slope 
profiles of halos of a given peak height, $\nu$, at $r \gtrsim R_{200}$ are remarkably similar 
at different redshifts when radii are scaled by $R_{200}$. 
This suggests that $R_{200}$ is suitable for describing the structure and evolution 
of the density profile in the outskirts of halos.

We then constructed the halo merger trees using the algorithm described 
in \cite{Springel2005Natur}. For a given halo `A' identified at a snapshot
`$i$', we track all particles of `A' in the next snapshot `$i+1$' and  
give a certain weight to each particle according to its binding energy.  
We identify halo 'B' as the descendant of halo `A' in snapshot `$i+1$'
if it has the highest score among all halos containing these particles. 
It is thus possible that one halo may have more than one progenitor,  
but a halo can only have one descendant. The branch that traces the 
main progenitor of a main halo back in time is referred to as the main 
trunk of the merger tree of the main halo. For a main halo at $z=0$, 
we estimate its formation redshift, $z_{\rm f}$, the highest redshift 
when the main trunk reaches half of its final halo mass.

In order to obtain sufficiently accurate density profiles, each halo needs to 
contain a sufficiently large number of particles. We used halos with more than 3000 particles, 
corresponding to $\log(M_{200}/\msun) > 12.0$. We only considered main halos 
because the density profiles of subhalos can be affected by the massive main 
halos. We divided the selected halo sample into five mass bins as shown 
in Fig. \ref{DM}. The number of halos in each mass bin is indicated in 
the corresponding panel in Fig.~\ref{DM}.

\subsection{The large-scale tidal field}
\label{sec_tid}

To explore the anisotropy in halo profiles, we adopt the `halo tidal field' proposed 
by \cite{WangH11} to define the local reference frame for a halo. 
The tidal field is obtained from the distribution of halos above a certain mass 
threshold $M_{\rm th}$, and can in principle be estimated from observation. 
As shown in \cite{Yang05,Yang07}, galaxy groups/clusters properly selected from 
large galaxy redshift surveys can be used to represent the dark halo population, 
and halos with masses $\log(M_{200}/\msun) \gtrsim 12$ can be identified 
reliably in the low-$z$ universe. We thus adopt $\log(M_{\rm th}/\msun) = 12$. 

The tidal field tensor at the location of a halo, $h$, can be written as \citep{Chen16}, 
\begin{equation}
\label{eq_tidal}
\begin{aligned}
\mathcal{T}_h&=\sum^{N}_{i=1}\frac{R_{i}^{3}}
{2r^{3}_{i}}(\overrightarrow{r_{i}}\overrightarrow{r_{i}}).
\end{aligned}
\end{equation}
Here $r_{i}$ is the distance from the $i$th halo with mass greater than 
$M_{\rm th}$ to halo `h', and $\overrightarrow{r_{i}}$ is the 
corresponding unit vector.  $R_{i}$ is the virial radius ($R_{200}$) of halo `$i$'. 
We only use halos with $r_{i}<100\mpc$ to calculate the tidal tensor, 
and $N$ is the number of halos used in the calculation. 

We can obtain three eigenvalues, $t_{1}$, $t_{2}$, and $t_{3}$ and the corresponding 
eigenvectors ($T_{1}$, $T_{2}$, $T_{3}$) by diagonalizing the tidal field tensor. 
These three eigenvalues are defined so that $t_1>t_2>t_3$, and by definition
$t_1+t_2+t_3=0$. The vectors $T_{1}$, $T_{2}$, and $T_{3}$ represent the major, 
intermediate and minor axes of the tidal field, respectively. Defined in this way, 
$T_{1}$ corresponds to the direction of stretching of the external tidal force, 
while $T_{3}$ corresponds to the direction of compression. Thus, nearby structures, 
such as filaments or massive halos, tend to reside along $T_1$. In contrast, 
$T_3$ tends to point towards low-density regions. 
As shown in our previous studies \citep[e.g.][]{WangH11, Chen16}, the tidal 
force, $t_1$, can be used to characterize the environment of halos. 
The tests presented in these studies showed that the principal axes of the halo 
tidal field are strongly aligned with those estimated from the mass density 
field \citep[i.e. the method presented in][]{Hahn2007MNRAS} and our method is 
also valid at high redshift. We refer the readers to \cite{WangH11} and 
\cite{Chen16} for detailed tests of the halo tidal field.

\subsection{Density-gradient profiles and the fitting method}
\label{sec_denp}

We first sample the density profile of each halo in 25 logarithmically spaced bins 
between $0.1 R_{200}$ and $10 R_{200}$. We then derive the mean density profile 
for a given halo sample, and obtain the density gradient profile using the 
mean density profile. More specifically, the gradient at each radius bin is estimated 
by using the densities in the two adjacent bins. The error bars of the density 
and gradient profiles are calculated using 1000 bootstrap samples. We take the 
16-84 percentile range as the $1\sigma$ error on our measurements. 

We consider two types of density and gradient profiles. The first is based on 
profiles averaged over all directions, as was adopted in many earlier studies.
We will refer to such a profile as the total profile. The second is based 
on the profiles along the major ($T_{1}$) and minor ($T_{3}$) principal axes 
of the local tidal tensor, respectively. These profiles are obtained by using 
dark matter particles within a cone around the corresponding principal axis
with an opening angle of 30 degrees. We also tested with other opening angles, 
and found that none of the results changes significantly as long as the opening 
angle is below 30 degrees. The profiles obtained this way are referred to as 
the $T_1$ and $T_3$ profiles, respectively. 

In order to determine the location of an extremum in a profile to obtain a 
characteristic scale, most of previous studies fitted the density and gradient 
profiles using a model proposed by \cite{Diemer14} (Hereafter the DK14 model). 
Here we use the Gaussian Process Regression (GPR) method as implemented in 
the PYTHON package SCIKIT-LEARN \citep{Pedregosa11} to fit the discrete data points. 
The GPR algorithm is one of the most widely used machine-learning algorithms in 
processing and analyzing data. Here we use GPR as a flexible, 
non-parametric method to fit profiles \citep[see also][]{Han2019MNRAS}.  
For more details about the method we refer the reader to \cite{Rasmussen05}.

More recently, \cite{O'Neil22} fitted the density and gradient profiles with 
both the DK14 model and the GPR method. They found that both methods produce 
reliable results, and that the D14 method performed better for  
gradient profiles. We performed tests using both methods to fit the two types 
of profiles, and found that the uncertainties in the DK14 model are larger because 
of the complexity of the profiles. In contrast, all the profiles can be well 
fitted by the GPR method, with locations of all significant extrema identified 
reliably (see below). More importantly, we found that some small 
local minima and peak features, which are of interest to us, cannot be 
identified by DK14 model. Because of these, we decided to adopt the GPR method.

To estimate the uncertainties in the characteristic scales, we chose to use two 
levels of bootstrap sampling technique. For any given halo sample, we first 
generated 100 level-one (L1) bootstrap samples. For each L1 sample, we generated 
1000 level-2 (L2) bootstrap samples. For any L1 sample, we used the 1000 L2 
samples to estimate the errors of the mean gradient profile. These errors 
are used in the GPR to fit the L1 gradient profile and to determine
local extrema (hence characteristic scales, see next section) 
from the best-fitting curve. Finally we estimated the errors of the characteristic 
scales using the 16-84 percentile of the 100 L1 samples. Similar techniques 
have been used in previous investigations \citep[e.g.,][]{O'Neil22}.

\begin{figure*}[htb]
    \centering
    \includegraphics[scale=0.40]{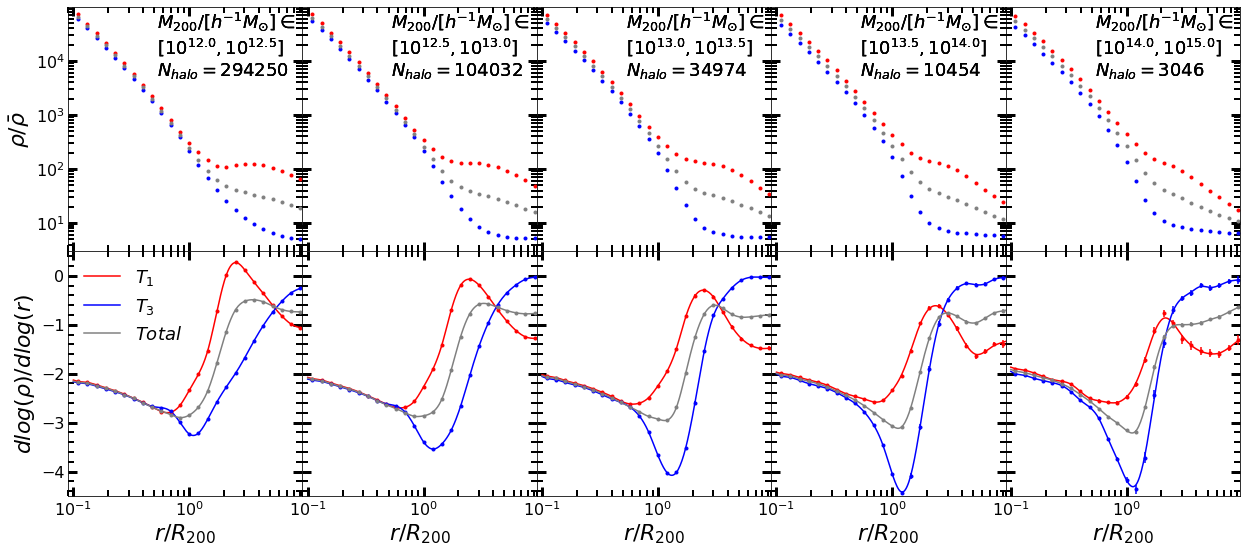}
        \caption{The data points show the mean dark matter density (upper panels) and density gradient (lower panels) profiles for halos in five mass bins. The solid lines represent the best-fitting profiles using GPR method. The gray symbols and lines show the results averaged over all directions, red for results along $T_1$ direction, and blue for $T_3$ direction. Error bars are estimated using 1000 bootstrap samples. 
        %{\bf 
        The halo mass ranges are shown in the upper panels.
        %} 
        The number of halos in each mass bin is shown in the corresponding upper panel. Please see Section \ref{sec_denp} for the details.}
        \label{DM}
\end{figure*}

\begin{figure*}[htb]
    \centering
    \includegraphics[scale=0.40]{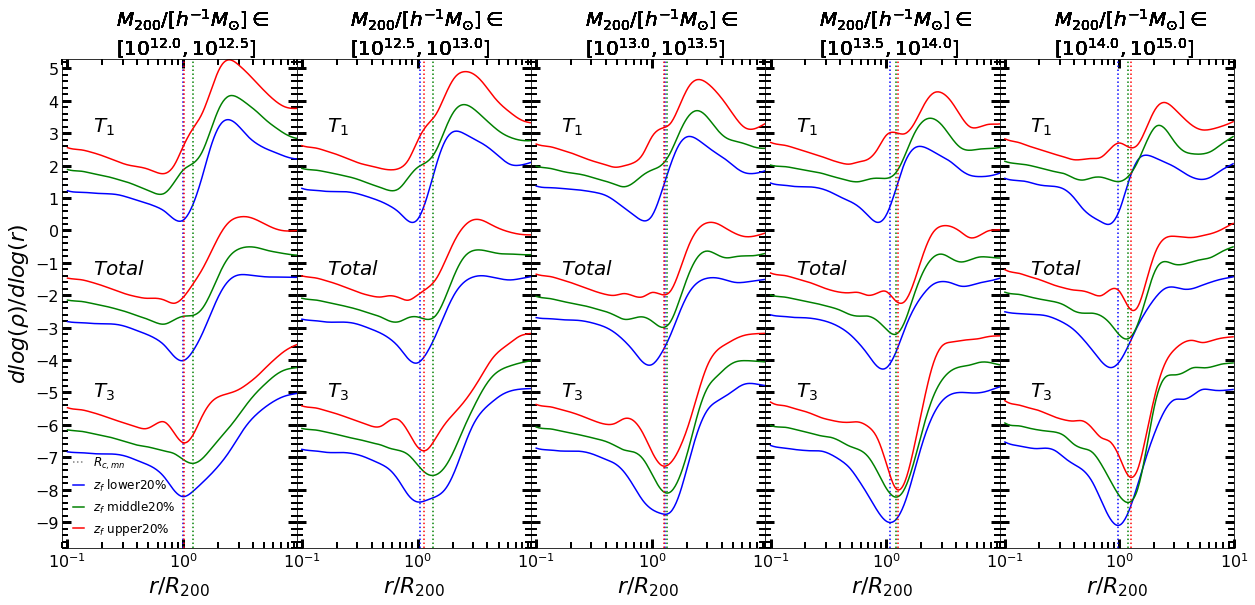}
        \caption{The best-fitting curves for gradient profiles. Different columns show the results of different halo mass, as indicated in each panel. The top, middle, and bottom sets of curves show the $T_1$, total and $T_3$ profiles, respectively. The profiles are shifted up or down for presentation purpose.  The red, green and blue lines represent halos in the upper, middle and lower 20\% of the $z_{\rm f}$ distribution in each mass bin. The vertical dotted lines indicate the splashback radii (the first caustic radii) determined from the $T_3$ profiles (See texts for details).}
        \label{fig_hprof_zf}
\end{figure*}

\begin{figure}[htb]
    \centering
    \includegraphics[scale=0.55]{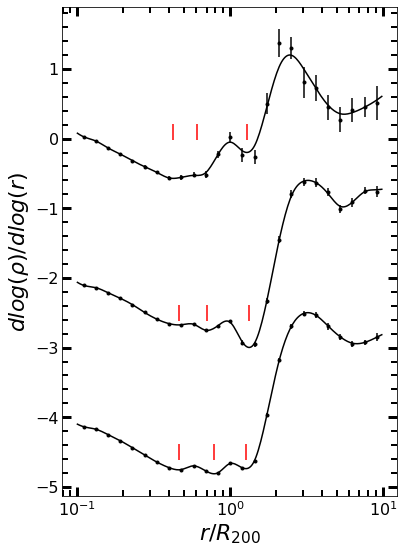}
        \caption{Three gradient profiles that have three local minimums at $0.25R_{200}<r<2R_{200}$. The points with error bars show the data and the solid lines show the best-fitting curves. The red vertical lines mark the locations of the measured local minimums.}
        \label{fig_check_three}
\end{figure}

\section{Characteristic scales in halo profiles}
\label{sec_iden}

In this section, we study in detail a number of characteristic scales present 
in the density gradient profiles. In Section \ref{sec_dm}, We show the density 
and density gradient profiles for halos of different masses and along 
the two principal axes of the tidal tensor. We use the GPR method to 
fit these profiles and identify the characteristic scales.
We investigate the dependence of  the characteristic scales on halo mass, 
formation time and environment in Sections \ref{sec_spdepend} and \ref{sec_pkdepend}. 
In Section \ref{sec_ps}, we examine the particle distribution in phase space
to understand how the characteristic scales in density profiles 
are linked to the dynamics of halo formation. In what follows, characteristic 
scales are usually expressed in term of the halo virial radius. 

\subsection{Halo profiles and characteristic scales}
\label{sec_dm}

Fig. \ref{DM} shows the mean mass density and density gradient profiles for 
five halo mass bins at z = 0. These profiles are the averages over all directions. 
In general, halos of different masses have similar density profiles from the inner 
region to several virial radii. At $r<R_{200}$, the density profiles can be well 
described by a universal formula, such as the 
Navarro–Frenk–White density profile \citep[e.g.][]{Navarro97}. 
At scales much larger the virial radius, the density profile is usually 
dominated by the two-halo term. From the gradient profile, we can see that the 
density gradient decreases with $r$ at $r\ls R_{200}$, and then increases rapidly 
up to $r= 2\sim3R_{200}$, remaining roughly at a constant value at larger $r$.
In the transition region around $r=R_{200}$, one can see a deep valley in the 
gradient profile, with a minimum (the steepest) gradient of about $-3$. This feature is 
often referred to as the ``splashback feature'', and the corresponding 
position is referred to as the ``splashback radius'' 
\citep[e.g.][]{Adhikari14,Diemer14,More15}. We can see that the splashback 
feature becomes deeper as the halo mass increases. This is 
consistent with previous results that this feature becomes more pronounced 
as the peak height, $\nu$, (equivalent to halo mass) increases 
\citep[][]{Diemer14,Diemer21}.

It is well known that both the density distribution and accretion 
around halos are anisotropic and both aligned with the large scale 
tidal field \citep[e.g.][]{Hahn2007MNRAS, WangH11, Shi15}. It is thus 
interesting to examine the density and gradient profiles along 
the major ($T_{1}$) and minor ($T_{3}$) principal axes of the tidal tensor 
(see Section \ref{sec_denp} for details). For comparison, the $T_{1}$ 
and $T_{3}$ profiles are also presented in Fig. \ref{DM}. 
As one can see, the density profiles change significantly with direction. 
Within $R_{200}$, the density along $T_{1}$ is usually higher than that 
along $T_{3}$, and the difference becomes larger as halo mass increases. 
This is consistent with the fact that the halo orientation is aligned with 
the large scale tidal field and the alignment becomes stronger with 
increasing halo mass \citep[e.g.,][]{Hahn2007MNRAS, ZhangY2009, WangH11,Tempel13,Libeskind13}. 
At $r\ge R_{200}$, the difference becomes even more prominent. 
For example, at $r\sim2R_{200}$, the density along $T_1$ is about 100 times 
the cosmic mean density, while that along $T_3$ is about 10 times lower. 
This may be expected as the $T_1$ vector is usually aligned with the 
large-scale filament containing the halo in question, and the $T_3$ vector 
usually points to low-density regions. Note that the density profile along 
$T_1$ at $r\sim2R_{200}$ is quite flat. 

In the density gradient profiles, we see two kinds of prominent features. 
The first one is a clear valley in both the $T_1$ and $T_3$ profiles, which is 
similar to the splashback feature in the total profiles.
The splashback feature along $T_3$ is deeper and narrower than that along $T_1$ 
in all halo mass bins, and the difference becomes larger as the halo mass increases. 
The location of the valley varies with direction, as to be quantified in 
Section \ref{sec_spdepend} below.
Recently, \cite{Contigiani21} examined a sample of massive halos, 
and found that the splashback feature is more prominent along the direction 
towards voids than that along filaments. This is consistent with our result, 
given the correlation between the local tidal field and the large-scale mass distribution. 
The second prominent feature is a peak around $2R_{200}$. This feature shows 
up in the gradient profiles along $T_1$, but is absent along the $T_3$ direction
and weak in the total profiles. Both the width and height of the peak decreases 
with increasing halo mass, and the peak gradient is around zero for low-mass halos. 

In order to quantify the characteristic features and scales, we need to measure 
the locations of the local extrema. For this purpose, we fitted the discrete density 
gradient profiles using the GPR method as mentioned in Section \ref{sec_denp}. 
The best-fitting results are shown in the lower panels of Fig. \ref{DM} as smooth curves. 
As one can see, the fits reproduce the gradient profiles very well. 
To check the quality of the fitting, we calculated the coefficient of 
determination, $R^2$, defined as 
\begin{equation}\label{eq_R2}
R^{2}=1-\frac{\sum_r[y(r)-y_{\rm GPR}(r)]^{2}}{\sum_r[y(r)-\bar{y}]^{2}},
\end{equation}
where $y(r)$ represents the data points as a function of $r$, $y_{\rm GPR}$ is the 
best-fitting curve, and $\bar{y}$ is the value averaged over all radius bins 
\citep[e.g.][]{Bucchianico08,Chicco21}. The closer $R^2$ is to 1, the better the fit. 
Our tests showed that all of the coefficients in our fitting are greater than 0.96,
indicating that the GPR method is able to catch significant features in the profiles.

The gradient profiles can become much more complicated than those shown in 
Fig. \ref{DM} when samples are divided according to the halo formation time.  
In our analyses, we divided the halo sample in a given mass range into five 
equal-sized sub-samples according to their $z_{\rm f}$. Fig. \ref{fig_hprof_zf} shows 
the best-fitting curves for halos in the upper, middle and lower 20\% 
of the $z_{\rm f}$ distribution. The $T_1$, $T_3$ and total profiles are all presented
for comparison. For clarity, we do not show the data points. As one can see, 
the peak feature in the gradient profile along $T_1$ is well defined in all cases,  
with its position showing some weak dependence on $z_{\rm f}$. 
The `valley' feature appears more complex and shows significant dependence 
on direction, halo mass and formation time. A number of local minima 
can be seen around the valley region for sub-samples of high $z_f$.
For example, one can see three local minima in the total profile of old halos 
with mass in the range between $10^{13}$ and $10^{13.5}\msun$.
To check whether or not this owes to potential over-fitting by the GPR, we show 
the data points for the profiles where three local minima are detected by GPR 
in Fig. \ref{fig_check_three}. These examples demonstrate clearly that 
the features are real and well captured by the GPR method. As discussed below, 
these features are likely caused by the caustics produced by particles at their 
(first or later) apocenter.

\begin{table*}[htb]
    \centering
    \caption{Classification of caustic radius and peak radius}
    \label{table1}    
    \begin{tabular}{c|c|c|c}
    \hline
         Classified by & Caustic Radius & Peak Radius & Note \\
         \hline
         \multirow{3}{*}{Direction} & $R_{\rm c,t}$ & $R_{\rm p,t}$ & Averaged over all directions \\
         %&\hline
                                    & $R_{\rm c,mj}$ & $R_{\rm p,mj}$ & Along $T_1$ (major axis) \\
                                    & $R_{\rm c,mn}$ & -- & Along $T_3$ (minor axis) \\
        \hline
        \multirow{3}{*}{\makecell[c]{Physical meaning \\ and location}} & First caustic & -- & First 
        apocenter (Splashback Radius), $>0.8R_{200}$ \\
                                          & Second caustic & -- & Second apocenter, $\sim0.6R_{200}$ \\
                                          & Third caustic & -- & Third or more apocenter, 
                                          $\sim0.4R_{200}$ \\
        \hline
    \end{tabular}
\end{table*}

With the help of GPR, we obtain the locations of the local extrema in the gradient 
profiles, corresponding to the characteristic scales that we are interested in. 
The location of a peak feature is represented by a peak radius, 
$R_{\rm p}$, and the location of a local minimum is denoted by a `caustic' radius, 
$R_{\rm c}$. In order to distinguish radii measured from different profiles, 
we use subscripts `t', `mj' and `mn' to denote measurements  
from the total, $T_1$ and $T_3$ profiles, respectively. For example, 
$R_{\rm c, mn}$ is the caustic radius measured from a $T_3$ profile while 
$R_{\rm p, mj}$ is the peak radius from a $T_1$ profile. Table~\ref{table1} lists 
all the measured radii and their definitions. Since peak features in 
$T_3$ profiles are weak or absent, we do not attempt to measure their locations.

As shown in Fig. \ref{fig_hprof_zf}, there are multiple local minima in the valley region.
To better understand the relation between the splashback radius and other 
caustic radius, we measure all the local minima in the valley region, 
over a radius range $0.25R_{200}<r<2R_{200}$. We find that each region can 
have at most three local minima. As shown below, the three caustics can be well 
separated according to their positions in $R_{\rm c}-z_{\rm f}$ space. 
We will refer to them as the first, the second and the third caustics, respectively, 
in the order of decreasing radius. As mentioned above, these caustics may be 
produced by particles at their first, second and third apocenters 
(see Section \ref{sec_spdepend}). 
This classification may thus have some bearing on physical processes 
underlying the generation of the caustics. The definitions
of these radii are also listed in Table~\ref{table1}. 

\subsection{Splashback, caustics and their dependencies on halo properties}
\label{sec_spdepend}

\begin{figure}[htb]
    \centering
    \includegraphics[scale=0.6]{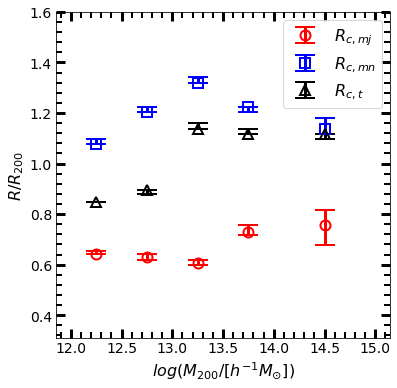}
        \caption{Caustic radius as a function of halo mass. The black, red and blue symbols show the results measured from total, $T_1$ and $T_3$ halo density gradient profiles.
        Error bars are calculated using the two-level (100$\times$1000) bootstrap samples(see Section \ref{sec_denp} for the details of the method). }
        \label{R_M}
\end{figure}

\begin{figure}[htb]
    \centering
    \includegraphics[scale=0.35]{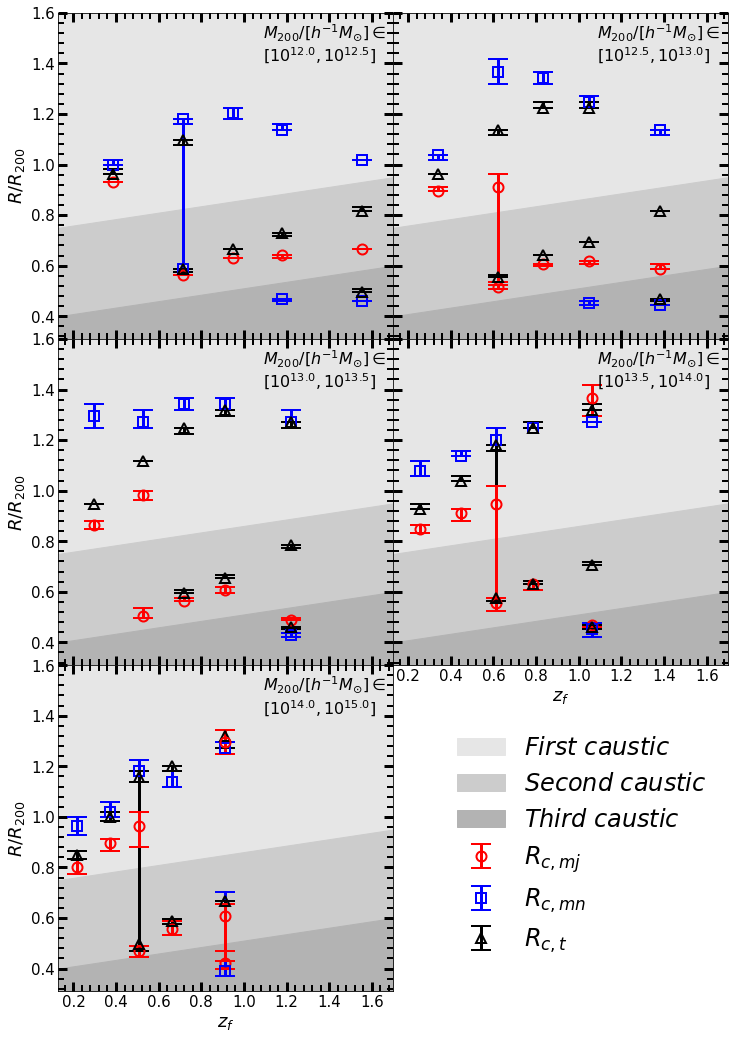}
        \caption{Caustic radii measured from total (black), $T_1$ (red) and $T_3$ (blue) profiles as functions of formation time in different halo mass bin, as indicated in each panel. Error bars are  calculated using the two-level (100$\times$1000) bootstrap samples. Note that some profiles have multiple local minimums, and thus multiple caustic radii. The shaded areas from light to dark indicate the regions, where the first caustic, the second caustic and the third caustic are located, respectively.}
        \label{R_zf_diff_M}
\end{figure}

\begin{figure}[htb]
    \centering
    \includegraphics[scale=0.35]{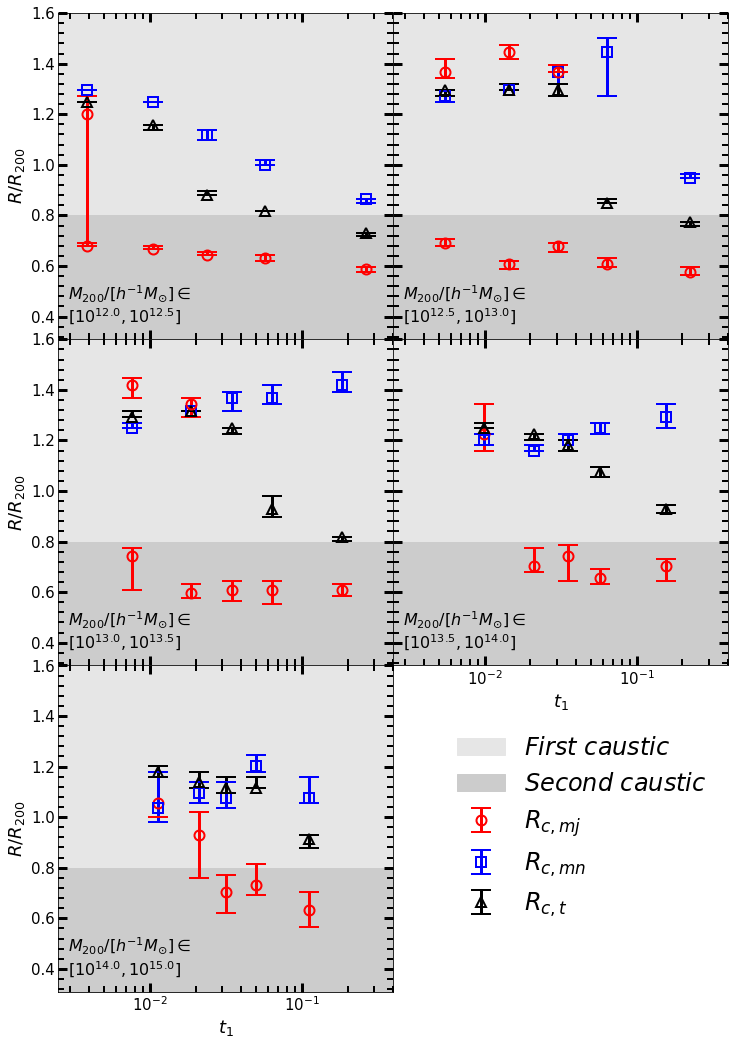}
        \caption{Caustic radii measured from total (black), $T_1$ (red) and $T_3$ (blue) profiles as functions of tidal force $t_1$ in different halo mass bin, as indicated in each panel. Error bars are  calculated using the two-level (100$\times$1000) bootstrap samples.The shaded areas from light to dark indicate the regions, where the first caustic and the second caustic are located, respectively.}
        \label{R_t1_diff_M}
\end{figure}

We first consider the dependence on halo mass. 
Fig. \ref{R_M} shows the caustic radius as a function of halo mass at $z=0$. 
As shown in Fig. \ref{DM}, there is only one local minimum in each gradient 
profile, so only one caustic radius is measured for each profile. 
As we can see, both $R_{\rm c,t}$ and $R_{\rm c,mn}$ first rise and then fall 
with increasing halo mass over the mass range in consideration. There seems to be 
a peak at $\log(M_{200}/\msun)\sim13$, which is broadly consistent with the results 
shown in figure 5 in \cite{Diemer14}. Some previous investigations focused 
on relatively massive halos, so their results can only show the monotonic 
decline with halo mass at the massive end \citep[e.g.,][]{O'Neil22}.  %{\color{red} Here, we should respond the comments of Diemer. In the interpretation of Figure 4, you say that some other studies had a limited mass range. That was true for O'Neil+22, but not for More+15 and my 2017/20 papers -- in the latter, the mass range goes down to ~1E10, so quite small. I think the correct explanation (to which you also allude in the text) is that the radius of steepest slope becomes less connected to splashback in these low-mass halos, and is simply an interplay between the orbiting (1-halo) and infalling profiles.}
In contrast, $R_{\rm c,mj}$ is almost independent of halo mass, 
and its value, which is in the range of $0.6\sim0.8R_{200}$,
 is less than that of $R_{\rm c,mn}$, which ranges from $1$ to $1.4R_{200}$. 
Previous studies also found anisotropy in the splashback/caustic radii measured 
with different methods \citep[e.g.][]{Mansfield17, Deason20}. However, 
as to be demonstrated in the following, the anisotropy  
can be explained by the difference between the first and second caustics, 
rather than the anisotropy in the splashback (first caustic).

Now we study the caustic radius as a function of halo formation time. 
As shown in Fig. \ref{fig_hprof_zf}, there are multiple local minima in the 
valley regions of some gradient profiles. For each sub-sample of formation time, 
we measure all the caustic radii, and the results are presented in 
Fig.~\ref{R_zf_diff_M}. The caustic radii appear to be in three distinct groups.
To show this more clearly, we split each panel into three regions, each 
containing data points in one group. The first group has $R_{\rm c,mj}$ 
and $R_{\rm c,mn}$ larger than $0.8R_{200}$, with typical values of about 
$1.1R_{200}$. The second group has caustic radius around $0.6R_{200}$, and the 
third one close to $0.4R_{200}$.  
The three groups are referred to as the first, second and third caustics, 
respectively, and mentioned above.
These three groups of caustics correspond to the 
three local minima shown in Fig. \ref{fig_hprof_zf}. 
Note that not all profiles have three measurable local minima,
because some caustics are weak and contaminated by the presence of 
nearby caustics. Indeed, a small fraction of measurements have large and 
asymmetric uncertainties, and some error bars are equal to the difference 
between two adjacent groups.
Note also that classification into the first, second and third 
caustics is different from that according to $R_{\rm c,t}$, $R_{\rm c,mj}$ 
and $R_{\rm c,mn}$. 

The first caustic appears in all $T_3$ profiles for all different $z_{\rm f}$, 
and can all be measured accurately, with results shown in Fig. \ref{R_zf_diff_M}. 
In contrast, only very young halos with $z_{\rm f}\ls0.6$ have measurable 
first caustic in their $T_1$ profiles. For the total profiles of massive halos
with $\log(M_{200}/\msun)>13$, the first caustic shows up for all $z_{\rm f}$ bins, 
while it is only detectable in the total profiles of lower-mass halos that have low $z_f$. 
As long as the feature is significant and the corresponding 
caustic radius can be measured, $R_{\rm c,mn}$, $R_{\rm c,mj}$ and $R_{\rm c,t}$ 
all follow a similar trend. They all increase with $z_{\rm f}$ at $z_{\rm f}<1$ 
and decrease with  $z_{\rm f}$ at $z_{\rm f}>1$. 
Previous studies \citep[e.g.,][]{More15,Diemer17,Diemer20b,Contigiani21,O'Neil21} 
found that the splashback radius decreases with increasing mass accretion 
rate for massive halos, which is consistent with the behavior seen here for 
the first caustic. The values of $R_{\rm c,mj}$ and $R_{\rm c,mn}$ 
for the first caustic are about the same for given $z_{\rm f}$ and halo mass. 
In most cases, $R_{\rm c,mn}$ is only slightly larger than $R_{\rm c,mj}$, 
suggesting that the first caustic is nearly isotropic. 

It is interesting to check how the first caustic looks like and whether 
it leaves any imprint in the $T_1$ profiles where it cannot be determined reliably.
In Fig. \ref{fig_hprof_zf}, we use vertical lines to mark the value of 
$R_{\rm c,mn}$ of the first caustic, corresponding to the steepest negative slope,  
measured from the gradient profiles along $T_3$. It can be seen that
the first caustic (if exists) does not always correspond to the lowest gradient value 
in the total and $T_1$ profiles. Interestingly, in the profiles where 
the corresponding local minimum does not exist, one still sees some signature 
of the first caustic around $R_{\rm c,mn}$, which is usually either weak 
or strongly affected by nearby features. These results suggest that
the first caustic actually exists along all directions for all halos, 
with a location quite independent of the direction defined by the tidal tensor. 

Next let us examine the second group of caustics. 
This feature is located around $0.6R_{200}$, very close to the 
location of the `second caustic' reported in previous studies 
\citep[e.g.,][]{Adhikari14, Diemer14, More15, Deason20, Xhakaj20}. 
This type of caustics tends to appear in relatively old halos, 
with $z_{\rm f}>0.5$, and is visible in the total and $T_1$ profiles 
but totally absent in $T_3$ profiles. As shown in Fig. \ref{fig_hprof_zf},
the second caustic is in general shallow and very close to the first one. 
In $T_3$ profiles this second caustic may be completely overwhelmed by 
the first, much deeper caustic, and thus difficult to measure.
As shown in Fig. \ref{fig_hprof_zf}, the second caustic radius 
exhibits a modest increase with $z_{\rm f}$. The third caustic resides in 
the more inner region of a halo, with $r\sim0.4R_{200}$. It is only visible 
in very old halos with $z_{\rm f}>0.9$, and is likely produced by particles 
that have completed more than two apocentric passages. Interestingly, 
this feature is more visible on the $T_3$ profiles than in the other two 
profiles (Fig. \ref{fig_hprof_zf}). In $T_3$ profiles, the third caustic 
is quite far from the first one, so that the contamination by the first one is reduced.

%{\bf 
\cite{Bertschinger85a} obtained an analytical solution for caustics
by assuming spherical symmetry and self-similar collapse for collisionless matter 
in an Einstein-de Sitter universe. They found that the radii of the first, second and third caustics 
are 0.364, 0.236 and 0.179 times the turnaround radius ($R_{\rm ta}$), respectively
\citep[see also][]{Mohayaee2006}. As shown in Section \ref{sec_ps}, $R_{\rm ta}$ is 
anisotropic and has a weak dependence on halo mass. The mean $R_{\rm ta}$ is about 
$2.5R_{200}$. Thus, the radii of the three caustics of the analytic solution in 
\cite{Bertschinger85a} correspond to 0.91, 0.59 and $0.45R_{200}$, respectively.
These are in good agreement with our finding for the local minima, providing 
additional evidence that these minima correspond to caustics in gravitational collapse.
%}

Fig. \ref{R_t1_diff_M} shows how the caustic radii depend on the strength 
of the tidal field, represented by the value of $t_1$, the eigenvalue of the 
tidal tensor along $T_1$. We divide the halo sample of a given mass bin into 
five equal-sized sub-samples in $t_1$, and present the caustic radii as functions 
of the mean $t_1$. There are sometimes multiple local minima in one profile, 
and we show the values for all the radii measurable.  
Consider first $R_{\rm c,mj}$ and $R_{\rm c,mn}$.
It is clear that there are two distinct groups of caustics. 
To guide the eye, we split each panel into two parts separated at 
$r=0.8R_{200}$. The caustics with radii larger than $0.8R_{200}$ 
correspond to the first caustic discussed above, and can be measured in 
all $T_3$ profiles as well as in some $T_1$ profiles in weak tidal 
fields. For halos of $\log(M_{200}/\msun)>13$, the caustic radius is totally 
independent of $t_1$. For low-mass halos, on the other hand, the radius 
decreases with increasing $t_1$. Halos with very large $t_1$ may 
reside in the splashback regions of nearby massive halos and their own 
splashback shells may have been affected severely by stripping effects.
The first caustic is nearly isotropic, with $R_{\rm c,mj}\approx R_{\rm c,mn}$ 
when both can be measured. 

The other group has caustic radii around 
$0.6R_{200}$, and thus corresponds to the second caustic. 
The second caustic can only be identified in $T_1$ profiles, and 
its radius is almost independent of $t_1$.
We are not able to identify the third caustic reliably, as it only appears 
in very old halos (Fig. \ref{R_zf_diff_M}) and likely is diluted by young halos.
In each of the total profiles, we can only identify one local minimum. 
At small $t_1$, the location of the minimum follows that of the first caustic. 
At large $t_1$, $R_{\rm c,t}$ lies between $R_{\rm c,mj}$ and $R_{\rm c,mn}$,
following the average between the first and second caustics. 
This indicates that the locations of the steepest slope in total profiles 
of halos with large $t_1$ are actually a mixture of those of the first and 
second caustics.

%{\bf 
Our analysis shows that the radius of the first caustic along $T_3$ depends on 
both formation time and environment.
Many factors, such as halo structure, halo merger, nearby structure (see next section), 
and velocity of the accreted material, can affect the first caustic and its measurement.
These factors are related to both formation time and environment, and thus can produce 
the dependencies we observe. This also suggests that the two dependencies may be related.  
As shown in \cite{WangH11}, $t_1$ is positively correlated with $z_{\rm f}$ for small halos. 
Thus, the decrease of the first caustic radius with $t_1$ may have the same origin as 
its decrease with $z_{\rm f}$ at $z_{\rm f}>1.0$ for small halos. 
At $z_{\rm f}<1.0$, the increasing trend with  $z_{\rm f}$ may be caused by the fact that
old halos are usually more compact than young ones. 
%}

The deep valley in $T_1$ and total profiles 
is sometimes a mix of several features. As an example, we reanalyze the 
results presented in Fig. \ref{R_M}. $R_{\rm c,mn}$ shown in the figure 
correspond to the first caustic, as it is always the dominant feature in 
$T_3$ profiles and its location depends only weakly on other properties. 
In contrast, the feature in $T_1$ profiles varies dramatically with $z_{\rm f}$. 
The first caustic dominates at small $z_{\rm f}$, while the second one dominates 
at large $z_{\rm f}$ (Figs. \ref{fig_hprof_zf} and \ref{R_zf_diff_M}). 
Consequently, $R_{\rm c,mj}$ shown in Fig. \ref{R_M}, an average over all 
$z_{\rm f}$, is the result of the blending of the two caustics. 
Being in the range of $0.6\sim0.8R_{200}$, the value of 
$R_{\rm c,mj}$ measured is closer to the second caustic. 
We thus conclude that the anisotropy shown in Fig. \ref{R_M} mainly 
reflects the difference between the first and second caustics, rather than 
the anisotropy of the splashback feature. In particular, the caustic radius in 
total profiles of low-mass halos is the mean result of the first and second caustics.
This is also true for the caustic radius obtained from the 
total profiles of halos with large $t_1$ (Fig. \ref{R_t1_diff_M}). 

The splashback radius studied in the literature corresponds to the first 
caustic, which is formed by particles at their first apocenter after infall 
\citep[e.g.][]{Diemer14,Adhikari14,More15,Shi16}. Our results demonstrate that 
the first caustic does not always correspond to the location of 
steepest slope or even a local minimum. Therefore, using the lowest point 
in the gradient profile to represent the location of splashback may lead to 
significant bias, as discussed above. Moreover, using the DK14 model to fit the 
density and gradient profiles for small and old halos or along $T_1$ 
may be inappropriate, as the profiles sometimes deviate significantly 
from the DK14 model (Fig. \ref{fig_hprof_zf}) \citep[see also][]{More15}. 
As shown above, the first caustic in $T_3$ profiles is prominent for both 
young and old halos over a large range in both halo mass and environment 
(represented by $t_1$). Our results also show that the location of 
the first caustic is approximately isotropic relative to the local tidal tensor.
Thus, the first caustic along $T_3$ may provide a promising way   
to study the splashback radius and its dependence on halo properties. 
As shown in \cite{WangH2012MNRAS}, the tidal field in the local Universe 
can be reconstructed from galaxy groups \citep{Yang07}. It is thus possible to 
have an unbiased measurement of the splashback radius from 
observational data by studying the galaxy distribution in the frame 
defined by the local tidal field.

\subsection{Gradient peak and its dependencies on halo properties}
\label{sec_pkdepend}

\begin{figure}[htb]
    \centering
    \includegraphics[scale=0.6]{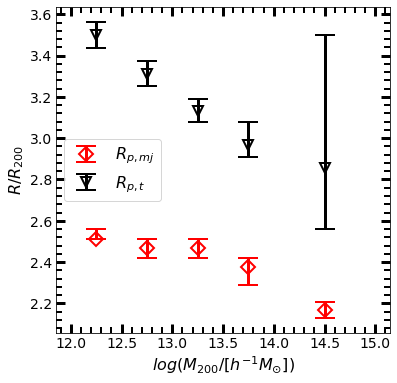}
        \caption{Peak radii measured from total (black) and $T_1$ (red) profiles as functions of halo mass. Error bars are calculated using the two-level (100$\times$1000) bootstrap samples.}
        \label{Rpk_M}
\end{figure}

\begin{figure}[htb]
    \centering
    \includegraphics[scale=0.6]{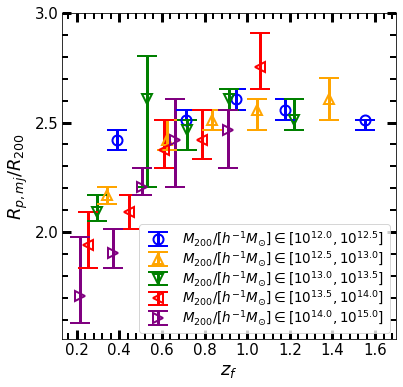}
        \caption{Peak radius along $T_1$ ($R_{\rm p,mj}$) as a function of formation time in different halo mass bin as indicated in the panel. Error bars are calculated using the two-level (100$\times$1000) bootstrap samples.}
        \label{Rpk_zf_diff_M}
\end{figure}

\begin{figure}[htb]
    \centering
    \includegraphics[scale=0.6]{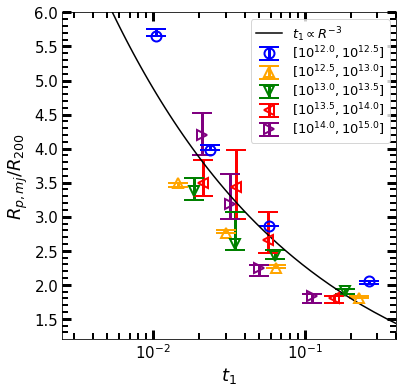}
        \caption{Peak radius along $T_1$ ($R_{\rm p,mj}$) as a function of the strength of tidal field ($t_1$) in different halo mass bin as indicated in the panel. Error bars are calculated using the two-level (100$\times$1000) bootstrap samples. The black curve shows $t_1\propto R_{\rm p, mj}^{-3}$ with an arbitrary amplitude.}
        \label{Rpk_t1_diff_M}
\end{figure}

Compared to the splashback and caustic features discussed above, 
the gradient peak feature is simpler: it appears as a single peak in the $T_1$ and total gradient 
profiles. We use the GPR method to fit the profiles and locate the local maxima to 
measure the peak radii. Fig. \ref{Rpk_M} shows $R_{\rm p,t}$ and $R_{\rm p,mj}$ 
as functions of halo mass. $R_{\rm p,t}$, ranging from 3.5 to 2.8$R_{200}$, 
declines quickly with halo mass, while $R_{\rm p,mj}$ is around $2.5R_{200}$, 
almost independent of halo mass except at the massive end. In all cases, 
$R_{\rm p,t}$ is larger than $R_{\rm p,mj}$. Examining the gradient profiles
in Fig.\ref{DM}, we can see that, around $R_{\rm p,mj}$, the 
gradient in the $T_3$ profiles is low but increases quickly with $r$, 
in contrast to the gradient in $T_1$ profiles, which starts to drop quickly with $r$. 
The peak in the total profiles is a combined effect of both the $T_1$ and $T_3$ 
profiles, but does not indicate the existence of a particular structure. 
Note that the uncertainty in $R_{\rm p,t}$ is quite large, particularly for 
the most massive halos, because the peak is rather weak.
In the following, we only present results for $R_{\rm p,mj}$.

The dependence of $R_{\rm p,mj}$ on halo formation time is shown in 
Fig. \ref{Rpk_zf_diff_M}.  We can see that the peak radius of halos 
with different masses follows almost the same trend with formation time, although 
the mean formation time ($\bar z_{\rm f}$) depends strongly on halo mass.
At $z_{\rm f}<0.6$ the peak radius increases with increasing formation redshift, 
while the dependence disappears at $z_{\rm f}>0.6$, with 
$R_{\rm p,mj}\sim 2.5R_{200}$. This result clearly suggests that the weak 
mass dependence is the secondary effect of the $z_{\rm f}$-dependence combined with
the $M_{200}-\bar z_{\rm f}$ relation. This interesting correlation may be 
related to the assembly bias that indicates the dependence of halo assembly on 
environment\citep{Gao2005, WangH2007}.

We also study the correlation of the peak radius with the tidal force ($t_1$) 
by dividing halos of a given mass range into five equal-sized sub-samples in $t_1$. 
Fig. \ref{Rpk_t1_diff_M} shows the peak radius as a function of the mean tidal force. 
There is a strong dependence of $R_{\rm p,mj}$ on $t_1$, with the peak radius deceasing 
rapidly with increasing tidal field strength. If the tidal force on a halo is 
dominated by another structure with distance $r$ to the halo, the tidal force 
is proportional to $r^{-3}$. As a reference, we show a curve of 
$t_1\propto R_{\rm p, mj}^{-3}$ with an arbitrary amplitude. As one can see, 
the data points follow the $t_1\propto r^{-3}$ relation well, suggesting that 
the presence of a locally dominating structure (LDS) at $r\sim R_{\rm p,mj}$ may 
be the main cause of the local tidal field of a halo. 
We do not find a peak within $10R_{200}$ for halos in the lowest $t_1$ bins, 
suggesting that these halos reside in an environment without a LDS. 

\begin{figure}[htb]
    \centering
    \includegraphics[scale=0.6]{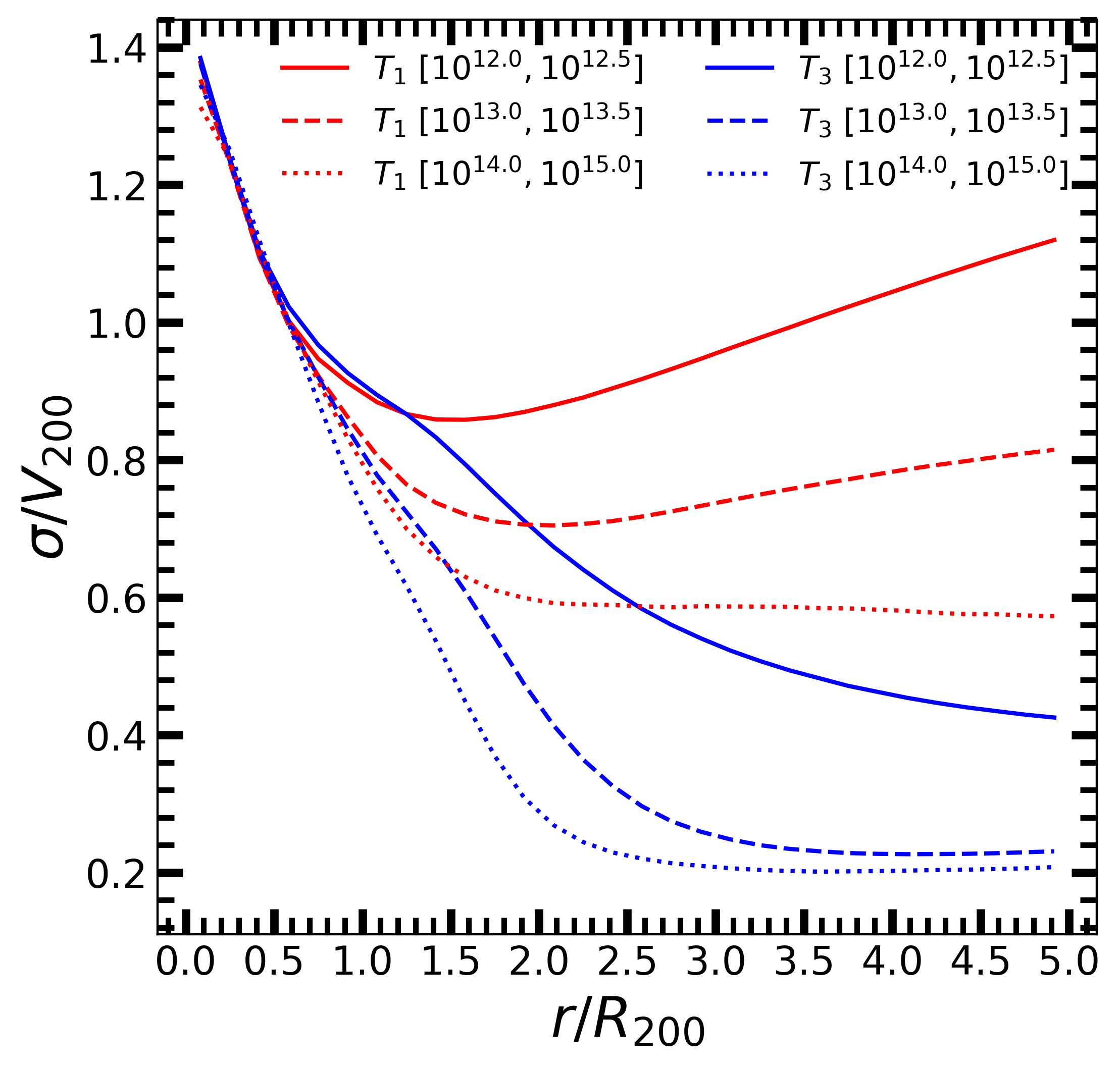}
        \caption{Velocity dispersion, $\sigma$, of dark matter particles as a function of radius. The velocity dispersion and radius are scaled with $V_{200}$ and $R_{200}$, respectively. Line styles denote different halo mass, as indicated. Red lines show the results along $T_1$ direction, and blue for $T_3$ direction.}
        \label{fig_sigma}
\end{figure}

\cite{Fong21} defined a characteristic depletion radius, $R_{\rm cd}$,  
using the location the lowest point of the halo bias profile
(defined as the ratio between the halo-mass cross correlation function and 
the mass auto-correlation function). 
They found that the value of $R_{\rm cd}$ is about $2.5R_{200}$, similar to the value
of $R_{\rm p,mj}$ we find here. They also found that
$R_{\rm cd}$, scaled with $R_{200}$, depends strongly on environment and only 
weakly on formation time and halo mass, similar to what we find for $R_{\rm p,mj}$. 
This may not be surprising, as the halo-mass cross correlation function is 
a measure of the mass density profile around halos. However, the bias 
function they used does not allow them to investigate any anisotropy in the mass 
distribution.

\cite{Anbajagane21} studied the thermal SZ effect in and around observed 
rich galaxy clusters, and found a clear flattening in the Compton-$y$ profile and
a prominent peak in the gradient profile at $\sim 2R_{200}$. 
The Compton $y$ parameter is the integration of the electron pressure 
along the line of sight, and thus is sensitive to both the gas density and gas 
temperature. In Fig. \ref{fig_sigma}, we show the velocity dispersion of dark 
matter particles as a function of radius along both $T_1$ and $T_3$ directions. 
At $2.5R_{200}$, the velocity dispersion of particles distributed 
along $T_1$ is much higher than that along $T_3$. For galaxy clusters, the difference 
is a factor of $\sim 3$, suggests that the baryonic gas in 
the $T_1$ direction is much hotter than that in the $T_3$ direction. 
From the density profile in Fig.\ref{DM}, one can also see 
that the mass density at $\sim 2.5R_{200}$ is about 10 times higher 
in the $T_1$ direction than in the $T_3$ direction.
All these suggest that the Compton $y$ parameter measured at such a radius
is dominated by the gas along the $T_1$ direction. 
Our results also indicate that this property of the SZ effect  
should also exist for low-mass halos, such as poor clusters and galaxy groups. 
The thermal SZ effect, therefore, provide a promising probe of the
peak feature in the mass distribution around halos.

Our results in Sections \ref{sec_dm} and \ref{sec_spdepend} 
show that (i) the density gradient profiles at $r<2R_{200}$ and caustics
along $T_1$ are very different from those along $T_3$ 
(Figs. \ref{DM} to \ref{R_t1_diff_M}); (ii) the 
first caustic along $T_1$ only appears in halos in weak tidal field 
(Fig. \ref{R_t1_diff_M}); (iii) the second caustic more likely appears 
along $T_1$ rather than along $T_3$ 
(Figs. \ref{R_zf_diff_M} and \ref{R_t1_diff_M}). 
All these suggest that the peak feature, i.e. the LDS, 
has a strong impact on the splashback and caustics. One possibility
is that the gravitational field of the LDS can change the halo 
accretion history. Indeed, as found in \cite{WangX21}, the assembly history of a halo 
is correlated with the density field outside the corresponding proto-halo 
in the initial density field. Another possibility is that the existence of 
the LDS affects the determination of the first caustic radius using 
the location of the local minimum. As shown in \ref{fig_hprof_zf}, 
the peak feature can sometimes extend to the region where the first caustic 
is expected to appear, particularly for low-mass halos and halos in regions of 
high-$t_1$ (see Figs. \ref{R_zf_diff_M} and \ref{R_t1_diff_M}).

\begin{figure*}[htb]
    \centering
    \includegraphics[scale=0.65]{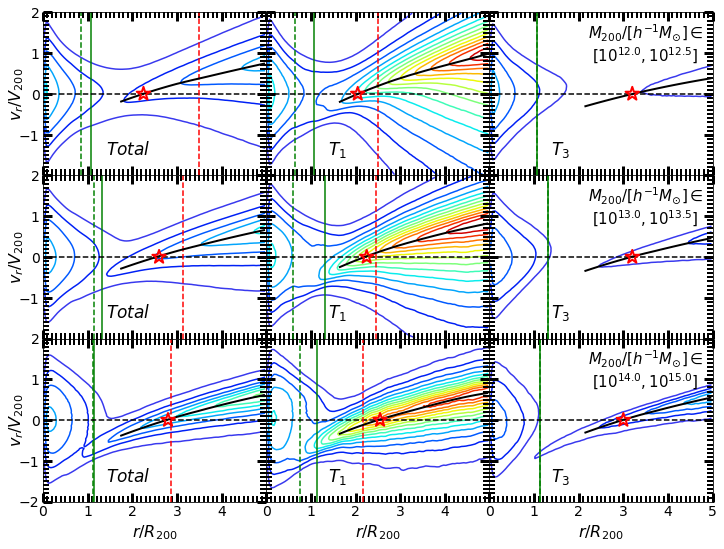}
        \caption{Phase space diagrams of dark matter particles in three halo mass bins. The left, middle and right columns show the results averaged over all direction (labelled as `total'), along $T_1$ and $T_3$, respectively. The contour lines are color coded by the phase space density(see eq. \ref{eq_ps}). Red(blue) means high(low) phase space density. The green and red vertical lines indicate the caustic and peak radii, respectively.
        The dashed vertical lines indicate the caustic radii measured in the total or $T_1$ profiles, while the solid vertical lines indicate the radii measured in the $T_3$ profiles and are also shown in total and $T_1$ panels.
        The black solid lines show the contour maxima of the accretion components. The red pentagram indicates the turnaround radius.}  
        \label{fig_phasespace}
\end{figure*}

\subsection{Signatures in phase space}
\label{sec_ps}

\begin{figure*}[htb]
    \centering
    \includegraphics[scale=0.50]{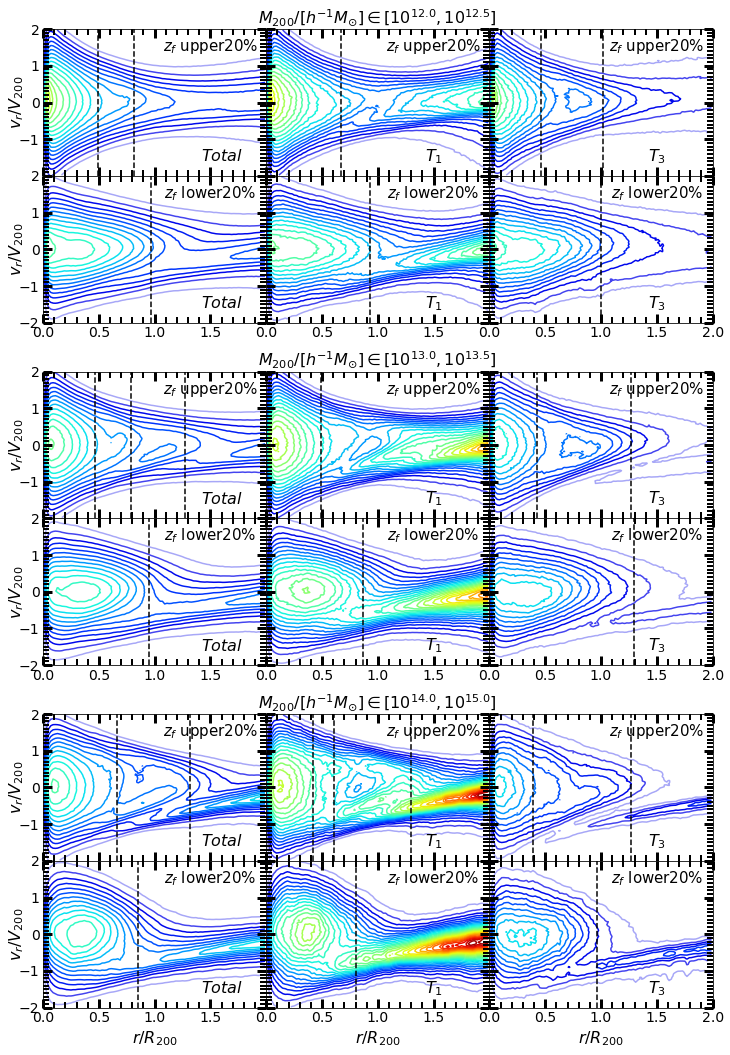}
        \caption{Phase space diagrams of dark matter particles for halos with different $z_{\rm f}$ and $M_{200}$. The panels are split into three groups according to the halo mass.
        In each group, the upper (lower) panels show the results for the oldest (youngest) 20\% halos, respectively.
        The left, middle and right columns show the results averaged over all direction (labelled as `total'), along $T_1$ and $T_3$, respectively. The contour lines are color coded by the phase space density(see eq. \ref{eq_ps}). Red(blue) means high(low) density. The black dashed lines indicate the caustic radii measured from gradient profiles.
        }
        \label{fig_phase_zf}
\end{figure*}

To understand the origin of the characteristic features, we examine
the dark matter particle distribution in phase space 
$(r/R_{200}, v_{\rm r}/V_{200})$. Here $r$ is the distance to the halo 
center and $v_{\rm r}$ is the radial velocity relative to the halo. 
The phase space density is expressed as,
\begin{equation}
\label{eq_ps}
\begin{aligned}
\rho_{\rm p}(r/R_{200}, v_{\rm r}/V_{200})=\frac{4\pi}{\Omega}\sum^{N}_{i=1}\frac{n_{i}(r/R_{200}, v_{\rm r}/V_{200})}{NM_{{200}, i}},
\end{aligned}
\end{equation}
where $N$ is the halo number in the halo sample, $M_{{200}, i}$ is the 
mass of halo $i$, and $n_{i}(r/R_{200}, v_{\rm r}/V_{200})$ is the number 
of dark matter particles associated with halo $i$ in the 
corresponding $r/R_{200}$ and $v_{\rm r}/V_{200}$ bin. The phase space density 
is normalized by the halo mass to eliminate potential mass dependence. 
Results obtained using all particles in all direction, $\Omega=4\pi$, 
are marked as `total' in Figs. \ref{fig_phasespace} and \ref{fig_phase_zf},  
while results along $T_1$ and $T_3$ use $\Omega=4\pi(1-\sqrt{3}/2)$, 
corresponding to a solid angle with an opening angle of 30 degrees. 
Note that the Hubble flow is included in the velocity, and a negative velocity 
indicates that the matter is falling onto the halo.

Fig. \ref{fig_phasespace} shows the phase space density distribution 
averaged over all directions (left), along $T_1$ (middle) and $T_3$ (right) 
for three halo mass bins. The results for the other two mass bins are similar
and not shown here. As one can see, there are two distinct components in the 
phase-space distribution. The first one has small $r$ and 
a broad and symmetric distribution in $v_{\rm r}$, which is composed 
of particles that are orbiting around the halo. We refer to this component as the 
halo component. The halo component has an extended tail outside the virial radius
with a small radial velocity. This is caused by splashback substructures 
together with diffuse mass \citep[see e.g.,][]{WangH09,Sugiura20,Diemer21}. 
The second component is mainly outside the virial radius, in which the mean 
radial velocity increases with distance. The inner part of this component 
penetrates into the halo component with an infall velocity, suggesting 
that halos are accreting material through this component. 
We refer to this component as the accretion component. \cite{Diemer21} separated 
particles according to orbital or infalling motions, and these two sets 
of particles are similar to the halo and accretion components defined here. 

As shown by red vertical lines in Fig.~\ref{fig_phasespace}, which 
mark the peak positions in the gradient profiles along $T_1$,  
the peak feature in the $T_1$ gradient profiles is associated with the 
accretion component. This component is much more prominent along $T_1$ than along $T_3$,
as $T_1$ tends to align with nearby filaments. The width of the velocity distribution 
for particles along $T_1$ is also much broader than that along $T_3$, and increases 
with decreasing halo mass, consistent with the velocity dispersion shown in 
Fig. \ref{fig_sigma}. Although our analysis above suggests the existence of  
LDS around $R_{\rm p,mj}$, we do not see any distinct structure in phase space. 
This is true even analyses are made for halos in regions of different $t_1$.
This suggests that the LDS along $T_1$ are mostly filamentary structures consisting
of different halos. 

The peak radius is close to the turnaround radius, $R_{\rm ta}$ (marked by red stars
in Fig.~\ref{fig_phasespace}), defined as the location where the mean radial velocity 
(marked with black solid lines) is equal to zero.
This suggests that the LDS responsible for the peak may be 
located at $\sim R_{\rm ta}$. Since the mean velocity is close to zero at $R_{\rm ta}$, 
mass has the tendency to be deposited at this radius, forming the peak feature
in the density gradient profile. However, along the $T_3$ direction, one does not 
see any significant signal in the density gradient profile at $\sim R_{\rm ta}$, 
because not much material is present to be accreted in this direction. 
We have constructed the phase-space maps for halos in regions of different 
$t_1$. Except for low-mass halos in the largest $t_1$ bin, $R_{\rm ta}$ 
along $T_1$ is almost independent of $t_1$, about $2\sim2.5R_{200}$, 
very different from the strong $t_1$-dependence of $R_{\rm p,mj}$ in this 
direction (see Fig. \ref{Rpk_t1_diff_M}). This indicates that $R_{\rm p,mj}$ 
is a characteristic radius very different from $R_{\rm ta}$.

 The green vertical lines mark the caustic radii defined using the gradient profiles. 
The values of $R_{\rm c,t}$ and $R_{\rm c,mj}$ are shown as the dashed lines 
only in their own panels, while for comparison $R_{\rm c,mn}$ is drawn 
as the solid lines in all the three panels (total, $T_1$ and $T_3$) for 
halos in a given mass bin. For the total sample, the halo and accretion components
are well separated by both $R_{\rm c,t}$ and $R_{\rm c,mn}$, 
as is expected from the similarity of the two radii (Fig. \ref{R_M}).  
For halos of the lowest mass, $R_{\rm c,mn}$ seems to perform better, 
even though it is defined from the $T_3$ profiles. For particles along the $T_1$ direction,
$R_{\rm c,mn}$ performs better than $R_{\rm c,mj}$ in all the three mass bins, 
consistent with the analysis in Section \ref{sec_spdepend}. We can see that 
the accretion component along $T_3$ is very weak, indicating that the $T_3$ profiles 
are dominated by the mass physically connected to halos, 
including mass both within the virial radius and the splashback mass outside the 
virial radius. Thus, $R_{\rm c,mn}$ may better represent the caustic formed by
particles at their first apocenters. Along $T_1$, the accretion component, 
in particular the peak feature, is prominent, making a significant contribution 
to the $T_1$ profiles around $R_{\rm c,mn}$ and affecting the assessment
of the splashback radius. 

It is also interesting to check the phase space density for halos with different 
$z_{\rm f}$. Fig. \ref{fig_phase_zf} shows the results for the 20\% oldest and 
the 20\% youngest halos in three mass bins. Here we want to examine inner regions of halos,
and so we only present particles within $2R_{200}$. For comparison, we mark the 
caustic radii determined above with vertical lines. For young halos, we can 
clearly see both the halo and accretion components in phase space, 
but no significant signal for the existence of sub-components is seen 
within the halo component. This is consistent with the fact
that only the splashback radii (first caustic) can be identified in the 
$T_1$ and $T_3$ profiles, as shown in Section \ref{sec_spdepend}. 
The splashback radius along $T_3$ is slightly larger than that along $T_1$.
This may indicate that the accretion flow along $T_1$, seen as the extension of the 
peak feature, can affect the splashback radius. Alternatively, the accretion component 
along $T_3$ may have larger infall velocities than that along $T_1$, and can thus reach 
to a larger apocenter.

Different from young halos, the old ones have clear sub-components within their 
halo components, so that multiple caustic radii can be identified. 
There seems to be one significant sub-component in the $T_3$ phase-space density distribution, 
which is seen as a symmetric excess in $\rho_{\rm p}$ between the two caustic radii 
and has $v_{\rm r}\sim 0$. This clump is very likely at its first apocenter, producing 
a prominent peak between the first and third caustics shown in the $T_3$ profiles 
of old halos (Fig. \ref{fig_hprof_zf}). One can also see an excess at the similar place
in the $T_1$ phase-space density distribution. The corresponding contour has a 
configuration aligned with the accretion component, suggesting that this excess 
is contaminated by the accretion flow. As shown in Fig. \ref{fig_phase_zf}, the caustic 
radii basically lie between the two (sub)components,  
and characterize the boundaries of the (sub)structures in halos. 

\section{Summary}
\label{sec_sum}

In this paper, we use the ELUCID N-body simulation to study the
characteristic features on the density gradient profiles in and 
around halos with masses from $10^{12}$ to $10^{15.0}\msun$. 
We use the GPR method to fit the density gradient profiles and find 
the locations of these characteristic features, represented by their scales.
We investigate the characteristic scales along the principal axes (the major axis 
$T_1$ and minor axis $T_3$) of the large-scale tidal field 
and their dependence on halo mass, halo formation redshift ($z_{\rm f}$) 
and environment (tidal field strength represented by $t_1$). 
We also verify our results and understand the origins of these features 
using the distribution of dark matter particles in phase space. 
Our main findings and conclusions are summarized below.

\begin{itemize}
\item In the density gradient profiles, there are two types of dominant 
features. One is a deep `valley', which corresponds to the caustic and splashback 
features, formed by dark matter particles at their apocenters after infall. 
The other is a prominent `peak' along the $T_1$ direction, produced by the mass 
distribution surrounding halos. 

\item The GPR method can fit the gradient profiles very well, and the performance is 
almost independent of the direction, halo mass, halo formation time and environment. 
It is able of detecting even weak structures in the gradient profiles.

\item The valley in the gradient profiles sometimes contains complicated sub-components. 
We identified three groups of local minima, referred to as the first, second and third 
caustics. Whether or not the three caustics appear depends on the direction
relative to the local tidal field, halo formation time and environment.

\item The first caustic radius corresponds to the steepest slope 
along the $T_3$ direction, and ranges from 0.8 to 1.4$R_{200}$. 
In gradient profiles along $T_1$ for old halos and for halos in 
strong tidal fields, the first caustic appears either as a local minimum 
(sometimes the deepest one) or as a small but significant change in the gradient. 
The radius of the first caustic along $T_3$ is slightly larger than that along $T_1$. 
The first caustic radius corresponds to the splashback radius, and is produced
by particles at their first apocenters after infall. 

\item The second caustic, around $0.6R_{200}$, can be identified along 
$T_1$ but not along $T_3$, and it appears in relatively old halos 
($z_{\rm f}>0.5$). The third caustic, around $0.4R_{200}$, is clearly present
in the $T_3$ profiles, and only appears in very old halos of $z_{\rm f}>0.9$. 
These two radii are produced by mass at the second and third apocenters.

\item 
%{\bf 
The radii of the first, second and third caustics are consistent with the prediction of self-similar gravitational collapse.
%}

\item Our analyses show that using the gradient profiles averaged over all directions 
or along $T_1$ may lead to a significant bias in estimating the splashback radius. 
It amplifies the anisotropy and may result in a biased estimation of the splashback radius. The splashback radius estimated from $T_3$ direction, on the other hand,
is reliable for both young and old halos and in both strong and weak tidal field.
Correcting for such bias, we find that the splashback radius is approximately 
isotropic around a halo.

\item The peaks in gradient profiles are around $2.5R_{200}$, which correspond to a 
significant flattening in density profiles.  The peak radius increases with 
$z_{\rm f}$ at $z_{\rm f}<0.6$ and becomes constant at $z_{\rm f}>0.6$. At 
fixed $z_{\rm f}$, it is independent of halo mass. 

\item Our results show that the peak feature is very likely produced by structures that 
dominate the tidal force.  The dominating structures have significant impact on the 
caustics, as indicated by the difference of the caustic structure seen 
between the $T_1$ and $T_3$ profiles.

\item 
The caustic radii are found to separate different (sub)components of the particle 
distribution in phase space. The first caustic radius, i.e. the splashback radius, 
obtained from the $T_3$ direction performs the best in separating the accretion flow 
from particles that are orbiting around halos. %The peak radius is, on average, close to the turnaround radius, at which the radial velocity is the smallest. {\color{red} but in the text we say they are different. Their values are also very different for different $z_f$}

\end{itemize}

Our results show that the splashback radius can be determined in an 
unbiased way, by measuring the gradient profile along the minor axis of the local 
large-scale structure. The measurement is reliable for both old and low-mass 
halos in various environments. The splashback radius so obtained can be used to study 
its correlations with halo properties and environments. We also investigated 
the connection between the caustics and the peak in gradient profiles and 
discussed how these two types of characteristic features are related 
to halo assembly in different environments. 
Our results show that the splashback feature along the $T_3$ direction 
is prominent, and may be detected in observational data 
%{\bf 
through galaxy correlation function \citep[e.g.][]{More16}, weak lensing \citep[e.g.][]{Gavazzi2006}, and dark matter annihilation \citep[e.g.][]{Mohayaee2006}. More tests are required to verify this. 
%}
Our results also suggest that 
thermal SZ effects are expected from the peak feature around halos of 
various masses 
%{\bf
\citep[see the SZ signal in][]{Anbajagane21}
%}
, in particular 
along the $T_1$ direction. All these predictions, together with the information 
provided by accurate reconstructions of the cosmic web, such as the ELUCID, can be tested 
using observational data. 
We will come back to some of these in the future.

\section*{Acknowledgements}
We thank the referee for a useful report.
This work is supported by the National Key R\&D Program of China (grant No. 2018YFA0404503), the National Natural Science Foundation of China (NSFC, Nos.  11733004, 12192224, 11890693, and 11421303), CAS Project for Young Scientists in Basic Research, Grant No. YSBR-062, and the Fundamental Research Funds for the Central Universities. The authors gratefully acknowledge the support of Cyrus Chun Ying Tang Foundations. We acknowledge the science research grants from the China Manned Space Project with No.
CMS-CSST-2021-A03. The work is also supported by the Supercomputer Center of University of Science and Technology of China.

\bibliography{ref.bib}

\bibliographystyle{aa}
\end{document}